\documentclass[a4paper]{JHEP3}
\usepackage{epsfig}

\usepackage{amsmath}
\usepackage{graphicx}

\title{\boldmath Non-dipolar Wilson links for transverse-momentum-dependent wave functions}

\author{Hsiang-nan Li$^{a,b,c}$,  Yu-Ming Wang$^{d,e}$
\footnote{SFB-CPP-14-82, TTK-14-31, TUM-HEP-964/14.}

{\it \small  $^a$Institute of Physics, Academia Sinica, Taipei, Taiwan 115, Republic of China}

{\it \small  $^b$ Department of Physics, National Cheng-Kung University, Tainan, Taiwan 701, Republic of China}

{\it \small  $^c$ Department of Physics, National Tsing-Hua University, Hsinchu, Taiwan 300, Republic of China}

{\it \small $^d$ Institut f\"{u}r Theoretische Teilchenphysik und Kosmologie RWTH Aachen, D-52056 Aachen, Germany}

{\it \small $^e$Physik Department T31, James-Franck-Stra${\ss}$e, Technische Universit\"{a}t M\"{u}nchen, D-85748 Garching, Germany}

}

\abstract{We propose a new definition of a transverse-momentum-dependent (TMD) wave function
with simpler soft subtraction for $k_T$ factorization of hard exclusive
processes. The un-subtracted wave function involves two pieces of non-light-like
Wilson links oriented in different directions, so that
the rapidity singularity appearing in usual $k_T$ factorization is regularized,
and the pinched singularity from Wilson-link self-energy corrections is alleviated
to a logarithmic one.
In particular no soft function is needed, when the two
pieces of Wilson links are orthogonal to each other. We show explicitly at one-loop
level that the simpler definition with the non-dipolar Wilson links exhibits the same
infrared behavior as the one with the dipolar Wilson links and
complicated soft subtraction. It is pointed out that both
definitions reduce to the naive TMD wave function as the
non-light-like Wilson links approach to the light cone. Their equivalence is then
extended to all orders by considering the
evolution in the Wilson-link rapidity.}


\begin{document}


\section{Introduction}

Light-cone wave functions are fundamental ingredients for the perturbative QCD factorization
of hard exclusive reactions. Apart from computing short-distance coefficient functions with
increasing accuracy in perturbation theory, advanced theoretical predictions for physical
observables cannot be achieved without deep understanding of nonperturbative hadronic wave
functions, which should be compatible with the factorization theorem and take on maximal
universality among different exclusive processes. Tremendous efforts have been devoted
to the understanding of collinear factorization properties for a large amount of hard
exclusive processes, such as the pion-photon transition form factor
\cite{delAguila:1981nk,Braaten:1982yp,Kadantseva:1985kb}, the pion electromagnetic
form factor \cite{FGO81,BT87,MNP99} and heavy-to-light transition form factors
\cite{Beneke:2000wa,Beneke:2004rc,Beneke:2005gs,Hill:2004if,Becher:2004kk}.
The corresponding light-cone distribution amplitudes are defined as non-local
matrix elements of light-ray operators with a rather intuitive Wilson-link structure.
Light-cone distribution amplitudes also serve as non-perturbative inputs in the
factorization formulas of correlation functions, which are used to construct
QCD light-cone sum rules for heavy-to-light transition form factors
\cite{Balitsky:1989ry,Belyaev:1993wp,Agaev:2010aq} and for hadron strong couplings
\cite{Belyaev:1994zk,Khodjamirian:2011jp}.

A transverse-momentum-dependent (TMD) wave function provides the three-dimensional
profile of the underlying structure of a hadronic bound state in the $k_T$ factorization
theorem. Compared to light-cone distribution amplitudes, it is nontrivial to establish a
well-defined TMD wave function as elaborated in \cite{Collins,Collins:2014loa}, in spite of many
phenomenologically successful applications of the $k_T$ factorization to hard
exclusive processes \cite{Nandi:2007qx,Li:2009pr,Li:2013xna,Li:2010nn,Li:2012nk}.
The point resides in the design of the associated Wilson links and the introduction
of soft subtraction, so that rapidity divergences \cite{Co03} and
Wilson-line self-energy divergences are avoided \cite{BBDM}. As
light-like Wilson lines are adopted in the un-subtracted TMD definition,
rapidity divergences from radiative gluons collimated to the Wilson lines are
produced \cite{CS82,Brodsky:2000ii,ICS08,Cherednikov:2008ua,Cherednikov:2009wk}.
As these rapidity divergences are regularized by rotating the Wilson lines away
from the light cone \cite{CS82} (a non-light-like axial gauge $n\cdot A = 0$ with $n^2\not=0$
was chosen actually), the self-energy divergences attributed to the infinitely long
dipolar Wilson lines \cite{BBDM} appear. To overcome the above difficulties,
complicated soft subtraction, which involves a square root of a ratio of
soft functions, has been suggested \cite{Collins:2011zzd}. This definition
is an improvement of the one with multiple non-light-like Wilson links in
\cite{JMY05} (see \cite{Aybat:2011zv} for an overview of TMD parton densities).
For comparison with the TMD parton densities defined in soft-collinear effective theory
\cite{Becher:2010tm,MP10,Mantry:2010bi,Echevarria:2012pw}, refer to \cite{Collins}.

In this paper we will propose a simpler definition for a TMD wave function, which
does not contain the square root of soft functions, but is compatible with the $k_T$
factorization theorem, namely, free of both rapidity and self-energy
divergences. The key is to rotate the Wilson links in the un-subtracted wave
function away from the light cone, and to orient the two pieces of non-light-like
Wilson links in different directions. The arguments to support this proposal
include: (i) the above rotation of the Wilson links serves as infrared regularization
for the rapidity and self-energy divergences; (ii) as long as collinear divergences are
concerned, the directions of Wilson links could be arbitrary; (iii) soft divergences
still cancel between the pair of diagrams, in which radiative gluons from the
Wilson links in arbitrary directions attach to the valence quark and to the
valence anti-quark, because of color transparency (or between virtual and real
corrections to an inclusive process); (iv) once the two pieces of Wilson links are
oriented in different directions, the dipolar structure is broken, and the
pinched singularity in Wilson-line self-energy corrections,
arising from the integrand $[(n\cdot l+i0)(n\cdot l-i0)]^{-1}$,
is alleviated into $[(n\cdot l+i0)(n'\cdot l-i0)]^{-1}$. The soft subtraction
required to remove this ordinary infrared singularity is much simpler.
We consider the special case with the two pieces of Wilson links being
orthogonal to each other, i.e., $n\cdot n'=0$ for demonstration, for which even no
soft function is needed.

In Sec.~\ref{section: dipolar Wilson lines}
we study the complicated definition of a TMD wave function with the dipolar Wilson
links \cite{Collins,Collins:2011zzd}, taking the pion wave function
extracted from the pion transition form factor as an example.
We discuss the essential difference between parton
densities for inclusive processes and wave functions for exclusive processes,
which concerns choices of the time-like or space-like gauge vector.
The novel definition for the TMD wave function with non-dipolar Wilson lines
is proposed in Sec.~\ref{section: non-dipolar Wilson lines}, whose infrared
behavior is explicitly shown to be the same as the complicated definition
at one loop. The equivalence between the simpler and complicated
definitions is extended to all orders by considering their
evolutions in the Wilson-link rapidity in Sec.~\ref{equiv}. We then conclude in
Sec.~\ref{section: summary} with a brief discussion on the extensions of our
proposals to the $B$-meson light-cone wave functions and polarized TMD parton
densities in spin physics.

\section{TMD wave function with dipolar Wilson lines}
\label{section: dipolar Wilson lines}

We consider the TMD pion wave function defined for the $k_T$ factorization of
the exclusive process $\gamma^{\ast} \to \pi \gamma$. The TMD pion
wave function constructed from the involved pion transition form factor
\cite{Nandi:2007qx,Li:2013xna}, following the suggestion of \cite{Co03},
is only free of rapidity divergences.
To remove both the rapidity and pinched singularities, the complicated soft
substraction factor with a square root \cite{Collins:2011zzd} is
introduced to the un-subtracted wave function:
\begin{eqnarray}
\phi^C(k_{+}^{\prime},k_T^{\prime},y_2)&=& \lim_{\substack{y_1 \to + \infty  \\   y_u \to - \infty}}  \,\,
\int \frac{d z_{-}}{2 \pi} \int \frac{d^2 z_{T}}{(2 \pi)^2}
\,  e^{i(k_{+}^{\prime} z_{-} -k_T^{\prime} z_T)} \, \nonumber \\
&& \times  \langle 0| \bar d(0) W^{\dag}_{u}(+\infty,0)  \! \not  n_{-} \, \gamma_5 \,
W_{u}(+\infty,z) \, u(z) |\pi^{+}(p)  \rangle \,
\nonumber \\
&&  \times   \sqrt{\frac{S(z; y_1, y_2)} {S(z;y_1,y_u) \, S(z;y_2,y_u)} } \,,
\label{TMD def: Collins}
\end{eqnarray}
with the coordinate $z=(0,z_{-},{\bf z}_T)$ of the $u$ quark field
and the Wilson link
\begin{eqnarray}
W_{n}(+\infty,z) =P \exp \left[ i g_s \int_0^{+\infty} \, d \lambda \, T^a \,
n \cdot A^a (\lambda n +z) \right] \,. \nonumber
\end{eqnarray}
The soft function in Eq.~(\ref{TMD def: Collins}) reads
\begin{eqnarray}
S(z;y_A,y_B) =\frac{1}{N_c} \langle 0|W^{\dag}_{n_B}(\infty,z)_{ca} \,
W_{n_A}(\infty,z)_{ad} \,  W_{n_B}(\infty,0)_{bc} \, W^{\dag}_{n_A}(\infty,0)_{db} | 0 \rangle,
\label{soft}
\end{eqnarray}
where $y_A$ and $y_B$ denote the rapidities of the gauge vectors $n_A$ and
$n_B$, respectively, and the color indices $a,b,c,d$ have been specified in
\cite{Collins:2011zzd}. The gauge vector $u$ associated with the un-subtracted wave
function approaches to the light-like direction $n_-=(0,1,{\bf 0}_T)$ in the limit
$y_u \to -\infty$. The vertical Wilson lines connecting
the longitudinal Wilson lines in Eq.~(\ref{TMD def: Collins}) at
infinity do not contribute in covariant gauge \cite{ICS08}.

In contrast to a space-like gauge vector for defining a TMD parton density in
Ref.~\cite{Collins:2011zzd,Collins:2004nx}, we have adopted the time-like
vector $n_2=(e^{y_2},e^{-y_2},{\bf 0}_T)$ with the rapidity $y_2$ in
the soft subtraction factor. Notice the essential difference between a
parton density and a wave function attributed to the final-state cut in the former.
The pinch singularity from the Wilson-line self-energy correction with a real
radiative gluon is only present in a TMD parton density with a space-like gauge vector,
but not in the one with a time-like gauge vector. As explained in \cite{BBDM}, the pole
of the involved eikonal propagator cannot be reached by an on-shell gluon under a
time-like gauge vector. However, the pinched singularity appears in the TMD wave
functions with both space-like and time-like gauge vectors, because the radiative
gluon is virtual. As indicated by the corresponding loop integrand
\begin{eqnarray}
\frac{n_2^2}{\left ( l^2 + i 0 \right ) \, \left ( n_2 \cdot l + i 0 \right )
 \left ( n_2 \cdot l - i 0 \right )}\,,
\end{eqnarray}
the minus component $l_{-}$ of the loop momentum is not bounded at all, so the
singularity at $n_2 \cdot l=0$ can be reached for a general $n_2$.
One can also find such a divergence from the loop integral
in coordinate space \cite{Collins}
\begin{eqnarray}
I &=& \int_0^{\infty} \, d \lambda_1 \int_0^{\infty} \, d \lambda_2 \,
\frac{1}{[(\lambda_1-\lambda_2) n_2 - z]^2}  \nonumber \\
&=&  \lim_{L \to \infty} \,\left [ {\pi \over \sqrt{z^2/n_2^2}} \, L
- \ln L +  \ln \left (\sqrt{{z^2 \over n_2^2 }}\right )  -1  + O(1/L) \right ]\,,
\end{eqnarray}
where the condition $n_2 \cdot z=0$ has been implemented to simplify the expression, and
$L$ denotes the length of the Wilson lines.

\begin{figure}[ht]
\begin{center}
\includegraphics[scale=0.8]{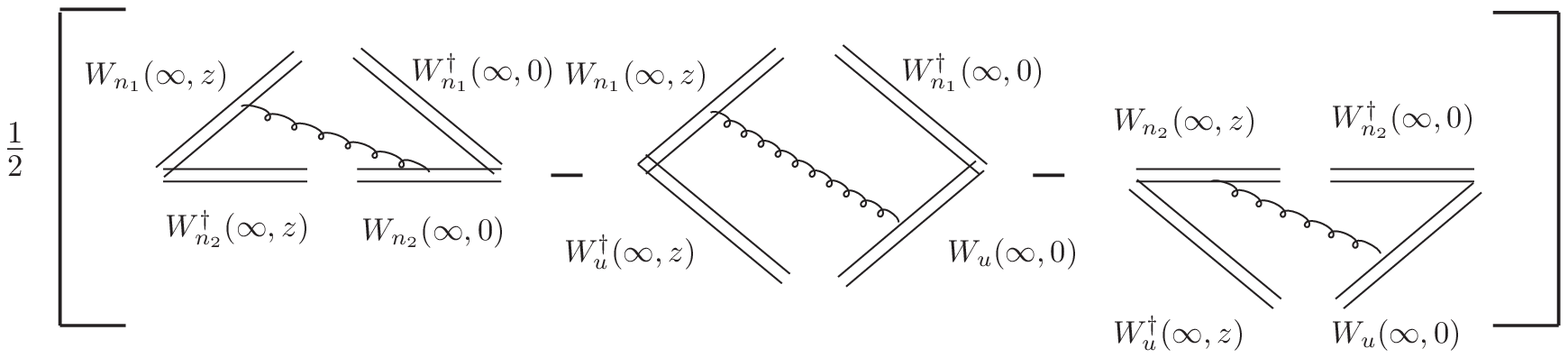}
\\(a)\\
\vspace{1 cm}
\includegraphics[scale=0.8]{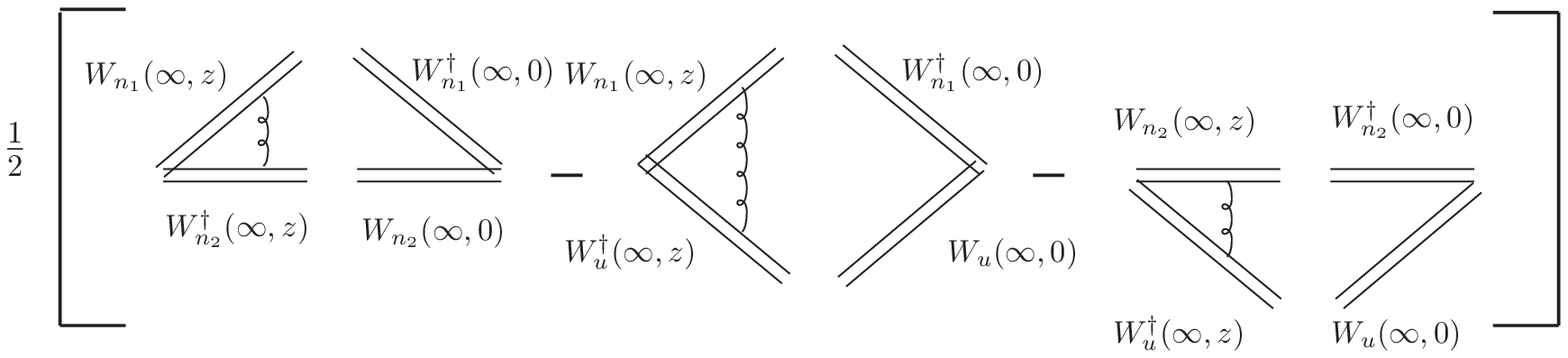}\\(b)
\caption{One-loop graphs for the soft subtraction factor
in Eq.~(2.1).}
\label{fig:soft_factor}
\end{center}
\end{figure}

It is a crucial criterion that the linear divergence proportional to the length
of Wilson lines should cancel in factorization-compatible definitions of a
TMD wave function, leading to one of the key requirements for
the construction of the soft subtraction factor.
The soft factor is designed in the way that the rapidity  divergences
associated with the gauge vector $n_1$ cancel between $S(z; y_1, y_2)$ and
$S(z;y_1,y_u)$, the pinched singularities in the self-energy corrections to the
Wilson lines in $n_2$, mentioned above, cancel between $S(z; y_1, y_2)$ and $S(z;y_2,y_u)$,
and the rapidity divergences in the un-subtracted wave function are cancelled by
$S(z;y_1,y_u)$ and $S(z;y_2,y_u)$ in the limit $y_u \to - \infty$.
These cancellations are easily understood from the typical one-loop diagrams for
the soft factor in Fig.~\ref{fig:soft_factor}. As to the order of taking limits of
various regulators, the prescription is as follows: (a) Take the trivial limit $L\to\infty$
for the length of the Wilson links; (b) Compute the un-subtracted wave
function and the soft functions in $D=4-2 \,\epsilon$ dimensions; (c) Take the limits of
infinite Wilson-line rapidities  $y_1 \to + \infty$ and $y_u \to - \infty$;
(d) Add the ultraviolet counterterms and remove the ultraviolet regulator by setting
$\epsilon \to 0$. Detailed discussions on the exchange
of the above limits can be found in Ref.~\cite{Collins:2011zzd}.

Figure~\ref{fig:soft_factor}(a) yields the integral
\begin{eqnarray}
S_a^{(1)}(k_{+}^{\prime},k_T^{\prime},y_2) = -g_s^2 \, C_F \, \mu^{2 \epsilon} \,
\int \frac{d l_{+}}{2 \pi} \int \frac{d^{2-2\epsilon} l_{T}}{(2 \pi)^{2-2\epsilon}} \, \, 
\, \delta(k_{+}^{\prime}-k_{+}-l_{+}) \, \delta ({\bf k}_T^{\prime}- {\bf k}_T- {\bf l}_T) \nonumber \\
\times \left [ \frac{\theta(l_{+}) \, \theta({\bar k}_{+}-l_{+})}
{l_{+}+i 0}  -  \frac{\theta(-l_{+}) \, \theta(l_{+}+k_{+}) } {l_{+}+i 0} \right ]
\frac{1} {l_T^2+ 2 \, e^{-2 y_2} \, l_{+}^2 +m_g^2- i 0 }  \,,
\label{soft factor a}
\end{eqnarray}
where $k_{+}^{(\prime)}$ and ${\bf k}_{T}^{(\prime)}$ denote the plus and transverse
components of the quark momentum before (after) the gluon emission for the partonic
configuration $|u(k) \, \bar d (p-k) \rangle$ in the Fock-state expansion of  $|\pi^+(p)  \rangle$,
and the shorthand notation ${\bar k}_{+}^{(\prime)}=p_{+}- k_{+}^{(\prime)}$ has
been employed\footnote{The primed components $k_{+}^{\prime}$ and ${\bf k}_{T}^{\prime}$
in the soft function $S_a^{(1)}(k_{+}^{\prime},k_T^{\prime},y_2)$ appear as the
conjugate variables to the coordinate $z$ in Eq.~(\ref{TMD def: Collins}) under the Fourier
transformation.}. The gluon mass $m_g$ regularizes the soft divergence to be
cancelled by the contribution from Fig.~\ref{fig:soft_factor}(b),
\begin{eqnarray}
S_b^{(1)}(k_{+}^{\prime},k_T^{\prime},y_2) =g_s^2 \, C_F \, \mu^{2 \epsilon} \,
\int \frac{d l_{+}}{2 \pi} \int \frac{d^{2-2\epsilon} l_{T}}{(2 \pi)^{2-2\epsilon}} \, \,
\delta(k_{+}-k_{+}^{\prime}) \, \delta ({\bf k}_T- {\bf k}_T^{\prime})  \nonumber \\
\times \left [ \frac{\theta(l_{+})} {l_{+}+i 0}  -  \frac{\theta(-l_{+})} {l_{+}+i 0} \right ]  \,
\frac{1} {l_T^2+ 2 \, e^{-2 y_2} \, l_{+}^2 +m_g^2- i 0 } \,.
\label{soft factor b}
\end{eqnarray}

\begin{figure}[ht]
\begin{center}
\hspace{-1 cm}
\includegraphics[scale=0.8]{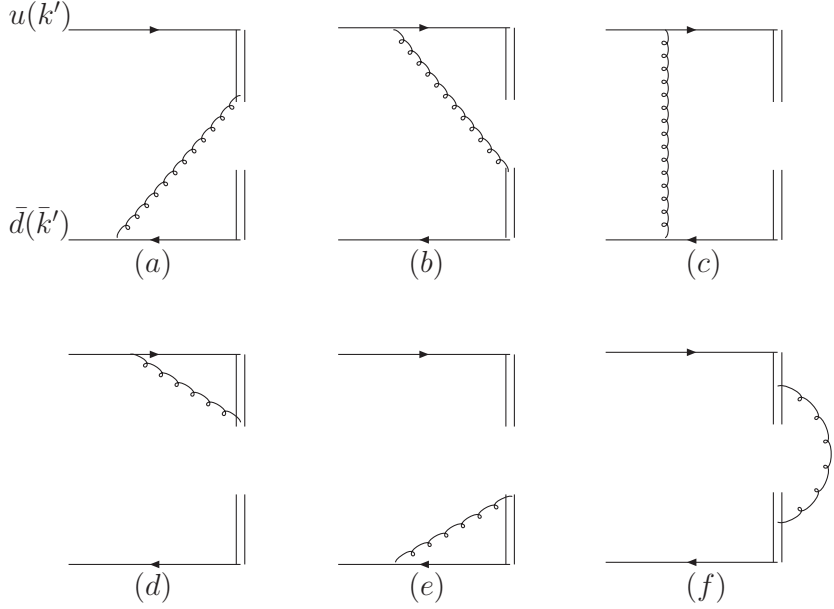}
\caption{One-loop graphs for  the un-subtracted TMD wave function.}
\label{fig:quark-wilson-diagram}
\end{center}
\end{figure}

The one-loop integrals for the un-subtracted TMD wave function from
Fig.~\ref{fig:quark-wilson-diagram} are written as
\begin{eqnarray}
\phi_a^{C(1)}(k_{+}^{\prime},k_T^{\prime}) &=& - g_s^2 \, C_F \, \mu^{2 \epsilon} \,
\int_{-{\bar k}_{+}}^0 \frac{d l_{+}}{2 \pi} \int \frac{d^{2-2\epsilon} l_{T}}{(2 \pi)^{2-2\epsilon}} \,
\delta(k_{+}^{\prime}-k_{+}+l_{+}) \, \delta ({\bf k}_T^{\prime}-{\bf k}_T+{\bf l}_T) \nonumber \\
&& \times \frac{1} {l_{+}+i 0} \, \frac{1}
{l_T^2- {l_{+} \over l_{+}+{\bar k}_{+}} \, \left({\bf l}_T- {\bf k}_T \right)^2 + i 0} \,,
\nonumber \\
\phi_c^{C(1)}(k_{+}^{\prime},k_T^{\prime}) &=& -i g_s^2 \, C_F \,  (2-2\epsilon) \, \mu^{2 \epsilon} \,
\int \frac{d^{4-2\epsilon} l}{(2 \pi)^{4-2\epsilon}} \,\,
\delta(k_{+}^{\prime}-k_{+}+l_{+}) \, \delta ({\bf k}_T^{\prime}-{\bf k}_T+{\bf l}_T) \nonumber \\
&& \times \frac{ ({\bf k}_T-{\bf l}_T)^2}{[(p-k+l)^2 + i 0]   [(k-l)^2 + i 0]   [ l^2 + i 0]} \,,\nonumber \\
\phi_e^{C(1)}(k_{+}^{\prime},k_T^{\prime}) &=&  g_s^2 \, C_F \, \mu^{2 \epsilon} \,
\int_{-{\bar k}_{+}}^0 \frac{d l_{+}}{2 \pi} \int \frac{d^{2-2\epsilon} l_{T}}{(2 \pi)^{2-2\epsilon}} \,
\delta(k_{+}-k_{+}^{\prime}) \, \delta ({\bf k_T}- {\bf k_T^{\prime}}) \nonumber \\
&& \times \frac{1} {l_{+}+i 0} \, \frac{1}
{l_T^2- {l_{+} \over l_{+}+{\bar k}_{+}} \,
\left({\bf l}_T- {\bf k}_T \right)^2 + i 0} \,, \nonumber \\
\phi_{b \, (d)}^{C(1)}(k_{+}^{\prime},k_T^{\prime}) &=& \phi_{a \, (e)}^{(1)}\,  \left [k_{+}^{(\prime)}  \to
{\bar k}_{+}^{(\prime)} \,, {\bf k}_T^{(\prime)} \to - {\bf k}_T^{(\prime)} \right ].   
\label{expression of fig. 2}
\end{eqnarray}
The contribution from Fig.~\ref{fig:quark-wilson-diagram}(f) vanishes in Feynman gauge
due to the light-like gauge link in the direction of $n_-$, and it is cancelled by
those of the corresponding diagrams from $S(z;y_1,y_u)$ and $S(z;y_2,y_u)$ in
arbitrary gauge as stated before.

To illustrate the cancellations of the rapidity singularities from $l_{+}=0$ and of the
pinched singularities from the Wilson-line self-energy corrections
to the pion wave function in Eq.~(\ref{TMD def: Collins}),
we present the explicit expression for the sum of $\phi_a^{C(1)}$,
$\phi_b^{C(1)}$, and $S_a^{(1)}$,
\begin{eqnarray}
\phi_a^{C(1)}+\phi_b^{C(1)}+S_a^{(1)}=
- \frac{\alpha_s \, C_F}{2 \pi} \, D_{red} \,\,
\delta(k_{+}-k_{+}^{\prime}) \, \delta ({\bf k}_T- {\bf k}_T^{\prime})
\nonumber \\
-\frac{\alpha_s \, C_F}{2 \pi^2} \,
\bigg \{ \frac{\theta(k_{+}^{\prime}-k_{+})\, \theta({\bar k}_{+}^{\prime})  }
{(k_{+}-k_{+}^{\prime})
\left [ \left({\bf k}_T^{\prime}- {\bf k}_T \right)^2
- { k_{+}-k_{+}^{\prime} \over p_{+}-k_{+}^{\prime}} {\bf k}_T^{\prime \,\, 2}  \right ] }
\nonumber \\
- \frac{\theta(k_{+}^{\prime}-k_{+})\, \theta({\bar k}_{+}^{\prime})  }
{(k_{+}-k_{+}^{\prime})
\left [ \left({\bf k}_T^{\prime}- {\bf k}_T \right)^2
+ 2 \, e^{-2 y_2} \, (k_{+}^{\prime} - k_{+})^2 \right ] }
\nonumber \\
+ \frac{\theta(k_{+}^{\prime})\, \theta(k_{+}-k_{+}^{\prime})  }
{(k_{+}^{\prime}-k_{+})
\left [ \left({\bf k}_T^{\prime}- {\bf k}_T \right)^2
- { k_{+}^{\prime} - k_{+} \over k_{+}^{\prime}} {\bf k}_T^{\prime \,\, 2}  \right ] }
\nonumber \\
-\frac{\theta(k_{+}^{\prime})\, \theta(k_{+}-k_{+}^{\prime})  }
{(k^{\prime}_{+}-k_{+})
\left [ \left({\bf k}_T^{\prime}- {\bf k}_T \right)^2
+ 2 \, e^{-2 y_2} \, (k_{+}^{\prime} - k_{+})^2 \right ] } \bigg \}_{\oplus}\,,
\label{result:sum of 2a, 2b and Sa}
\end{eqnarray}
with the factor
\begin{eqnarray}
D_{red}&=&2 \left (\frac {1}{\hat \epsilon} + \ln {\mu^2 \over k_T^2}  \right )
+ \ln^2 \left ({k_{+} \over m_g} \right )
+ \ln^2 \left ({\bar k_{+} \over m_g} \right ) 
+ \ln \left ( {k_T^2 \over m_g^2}  \right )
\cdot \ln  \left ( {k_T^2 \over  k_{+} \bar k_{+} }  \right )
\nonumber \\
&&- \frac{1}{2} \ln^2 \left (2 \, e^{-2 y_2} \right )  - \ln \left (2 \, e^{-2 y_2} \right ) \cdot
\ln \left ( {k_{+} \bar k_{+} \over m_g^2} \right )  +4 -{\pi^2 \over 6} \,,
\end{eqnarray}
and $1/\hat \epsilon \equiv 1/\epsilon -\gamma_E + \ln(4 \pi)$.
The ``$\oplus$" subtraction is defined as
\begin{eqnarray}
\left [ f(k_{+},k^{\prime}_{+},{\bf k}_T,{\bf k}_T^{\prime}) \right ]_{\oplus}
=f(k_{+},k^{\prime}_{+},{\bf k}_T,{\bf k}_T^{\prime}) - \delta(k_{+}-k_{+}^{\prime}) \,
\delta ({\bf k}_T- {\bf k}_T^{\prime}) \nonumber \\
\, \times
\int_{-\infty}^{+\infty} d q_{+} \int_{-\infty}^{+\infty} d^{2-2\epsilon} q_{T} \,
f(k_{+},q_{+},{\bf k}_T,{\bf q}_T) \,.
\end{eqnarray}
It is evident that Eq. (\ref{result:sum of 2a, 2b and Sa}) is free of the
rapidity divergence from $k_{+}=k^{\prime}_{+}$, and contains only the ordinary
logarithmic soft divergence regularized by the gluon mass. This logarithmic divergence
is cancelled precisely by that in the sum of $\phi_d^{C(1)}$,
$\phi_e^{C(1)}$, and $S_b^{(1)}$,
\begin{eqnarray}
\phi_d^{C(1)}+\phi_e^{C(1)}+S_b^{(1)} &=& \frac{\alpha_s \, C_F}{2 \pi}
\bigg \{ D_{red}  \, + {\Gamma(\epsilon)  \over 2 \, \epsilon} \, 
\left(2 \, e^{-2 y_2}\right)^{-\epsilon}
\left [ \left ( \frac{4 \, \pi \, \mu^2}{k_{+}^2} \right )^{\epsilon}
+  \left ( \frac{4 \, \pi \, \mu^2}{ \bar k^{2}_{+} } \right )^{\epsilon} \right ] \bigg \}
\nonumber \\
&& \times  \delta(k_{+}-k_{+}^{\prime}) \, \delta ({\bf k}_T- {\bf k}_T^{\prime}) \,.
\label{result:sum of 2d, 2e and Sb}
\end{eqnarray}
Evaluation of Fig.~\ref{fig:quark-wilson-diagram}(c) gives
\begin{eqnarray}
\phi_c^{C(1)}(k_{+}^{\prime},k_T^{\prime})&=&-\frac{\alpha_s \, C_F}{4 \pi}
\left [ \frac {1}{\hat \epsilon} + \ln {\mu^2 \over k_T^{2}}  + 1 \right ]
\delta(k_{+}-k_{+}^{\prime}) \, \delta ({\bf k}_T- {\bf k}_T^{\prime})
-\frac{\alpha_s \, C_F}{2 \pi^2} \,  {1 \over p_{+}} \, \nonumber \\
&&   \times \bigg \{  \frac{\theta(k_{+}^{\prime})\, \theta(k_{+}-k_{+}^{\prime})  }
{ \left({\bf k}_T^{\prime}- {\bf k}_T \right)^2
+ { k_{+}-k_{+}^{\prime} \over k_{+}^{\prime}} {\bf k}_T^{\prime \,\, 2}   } \,
+ \frac{ \theta(k_{+}^{\prime}-k_{+}) \, \theta({\bar k}_{+}^{\prime}) }
{ \left({\bf k}_T^{\prime}- {\bf k}_T \right)^2
+ { k^{\prime}_{+}-k_{+} \over p_{+}-k_{+}^{\prime}} {\bf k}_T^{\prime \,\, 2} }  \bigg \}_{\oplus} \,,
\label{result:sum of 2c}
\end{eqnarray}
which does not contain a soft divergence.

We then obtain the next-to-leading-order (NLO) TMD pion wave function
from Fig.~\ref{fig:soft_factor} and Fig.~\ref{fig:quark-wilson-diagram},
\begin{eqnarray}
\phi^{C(1)}(k_{+}^{\prime},k_T^{\prime},y_2)  &=&-
\frac{\alpha_s \, C_F}{4 \pi}
\bigg \{\frac {1}{\hat \epsilon} + \ln {\mu^2 \over k_T^{2}}  + 1 
-{\Gamma(\epsilon)  \over \, \epsilon}  \left(2 \, e^{-2 y_2}\right)^{-\epsilon}
\left [ \left ( \frac{4 \, \pi \, \mu^2}{k_{+}^2} \right )^{\epsilon}
+  \left ( \frac{4 \, \pi \, \mu^2}{ \bar k^{2}_{+} } \right )^{\epsilon} \right ] \bigg \}
\nonumber\\
&& \times\delta(k_{+}-k_{+}^{\prime}) \, \delta ({\bf k}_T- {\bf k}_T^{\prime})
-\frac{\alpha_s \, C_F}{2 \pi^2} \, \bigg \{ \frac{\theta(k_{+}^{\prime}-k_{+})\, \theta({\bar k}_{+}^{\prime})  }
{(k_{+}-k_{+}^{\prime})
\left [ \left({\bf k}_T^{\prime}- {\bf k}_T \right)^2
- { k_{+}-k_{+}^{\prime} \over p_{+}-k_{+}^{\prime}} {\bf k}_T^{\prime \,\, 2}  \right ] }
\nonumber \\
&& - \frac{\theta(k_{+}^{\prime}-k_{+})\, \theta({\bar k}_{+}^{\prime})  }
{(k_{+}-k_{+}^{\prime})
\left [ \left({\bf k}_T^{\prime}- {\bf k}_T \right)^2
+ 2 \, e^{-2 y_2} \, (k_{+}^{\prime} - k_{+})^2 \right ] }
\nonumber \\
&& + \frac{\theta(k_{+}^{\prime})\, \theta(k_{+}-k_{+}^{\prime})  }
{(k_{+}^{\prime}-k_{+})
\left [ \left({\bf k}_T^{\prime}- {\bf k}_T \right)^2
- { k_{+}^{\prime} - k_{+} \over k_{+}^{\prime}} {\bf k}_T^{\prime \,\, 2}  \right ] }
\nonumber \\
&& -\frac{\theta(k_{+}^{\prime})\, \theta(k_{+}-k_{+}^{\prime})  }
{(k^{\prime}_{+}-k_{+})
\left [ \left({\bf k}_T^{\prime}- {\bf k}_T \right)^2
+ 2 \, e^{-2 y_2} \, (k_{+}^{\prime} - k_{+})^2 \right ] }
\nonumber\\
&& +\frac{\theta(k_{+}^{\prime})\, \theta(k_{+}-k_{+}^{\prime})  }
{ p_{+}\left[\left({\bf k}_T^{\prime}- {\bf k}_T \right)^2
+ { k_{+}-k_{+}^{\prime} \over k_{+}^{\prime}} {\bf k}_T^{\prime \,\, 2}\right]   }
+ \frac{ \theta(k_{+}^{\prime}-k_{+}) \, \theta({\bar k}_{+}^{\prime}) }
{ p_{+}\left[\left({\bf k}_T^{\prime}- {\bf k}_T \right)^2
+ { k^{\prime}_{+}-k_{+} \over p_{+}-k_{+}^{\prime}} {\bf k}_T^{\prime \,\, 2} \right]}
\bigg \}_{\oplus}\,, \nonumber \\
\label{NLO wf tot: collins}
\end{eqnarray}
indicating that the remaining infrared divergence
in the NLO pion wave function is the collinear one regularized by the parton virtuality
$k_T^2$. To validate the $k_T$ factorization theorem for the pion transition form factor,
we show the infrared finiteness of the hard kernel obtained from matching the QCD
diagrams onto the effective diagrams $\phi^{C(1)}$.
The collinear logarithm $\ln k_T^2$ is extracted explicitly from the convolution of the NLO
pion wave function with the leading-order hard kernel $H^{(0)}$ of the
pion transition form factor:
\begin{eqnarray}
\int_{-\infty}^{+\infty} d k_{+}^{\prime} \int_{-\infty}^{+\infty} d^{2-2\epsilon} k_{T}^{\prime} \,\,
\phi^{C(1)}(k_{+}^{\prime},k_T^{\prime},y_2) \, H^{(0)}(k_{+}^{\prime},k_T^{\prime}) \nonumber \\
=-{\alpha_s \, C_F \over 4 \pi} \, \left [  \ln \left ( {k_{+} \over p_{+}} \right ) + 2 \right ] \,
\ln k_T^2  \, H^{(0)}(k_{+},k_T) +\cdots\,,
\label{substraction: Collins}
\end{eqnarray}
where the ellipsis represents the terms independent of $\ln k_T^2$ at leading power.
It is indeed the case that Eq.~(\ref{substraction: Collins}) cancels the $\ln k_T^2$ term
in the one-loop QCD diagrams for the pion transition form factor
given by Eq.~(20) of \cite{Nandi:2007qx}, as those from the self-energy
corrections to the external quarks are excluded.

\section{TMD wave functions with non-dipolar Wilson lines}
\label{section: non-dipolar Wilson lines}

In view of the complicated structure of the soft subtraction in Eq. (\ref{TMD def: Collins}),
it is in demand to construct factorization-compatible definitions of a TMD wave
function with simper subtraction factors for practical calculations.
We start with the un-subtracted TMD wave function in Eq.~(\ref{TMD def: Collins}),
where the future-pointing or past-pointing light-like Wilson links have been
appropriately chosen to facilitate the $k_T$ factorization by avoiding the Glauber region.
Certainly, the Glauber region does not exist in a simple process \cite{CL09,Chang:2009bk}
like the pion transition form factor considered here.
The Wilson links are then rotated away from the light cone, as done in
\cite{CS82,Collins:2011zzd}, to regularize the rapidity divergence. The key
of our proposal is that the two pieces of Wilson links are rotated into
different directions, such that the pinched singularity in Wilson-line self-energy
corrections, arising from the integrand $[(n\cdot l+i0)(n\cdot l-i0)]^{-1}$,
is alleviated into $[(n\cdot l+i0)(n'\cdot l-i0)]^{-1}$. Hence, the
soft subtraction required to remove this ordinary infrared singularity
with the non-dipolar Wilson links is simpler. The technique of rotating
the Wilson links has been also employed to derive various resummations for a
TMD wave function \cite{Li:2013ela}. We then need to examine whether
the above rotation of Wilson links would change the collinear logarithms
$\ln k_T^2$, which have been absorbed into the un-subtracted TMD wave function.
As postulated in the Introduction and demonstrated by explicit calculations below,
the new definition reproduces the correct collinear logarithms.

We consider the case with two orthogonal pieces of off-light-cone Wilson links:
\begin{eqnarray}
\phi^W(k_{+}^{\prime},k_T^{\prime},y_2)=\int \frac{d z_{-}}{2 \pi} \int \frac{d^2 z_{T}}{(2 \pi)^2} \,
e^{i(k_{+}^{\prime} z_{-} -k_T^{\prime} z_T)} \, \nonumber
\\
\times \langle 0| \bar d(0) W^{\dag}_{n_2}(+\infty,0)  \! \not  n_{-} \, \gamma_5 \,
W_{v}(\infty,z) \, u(z) |\pi^{+}(p)  \rangle \,,
\label{TMD def: Without soft factor}
\end{eqnarray}
where the gauge vectors $n_2$ and $v=(-e^{y_2},e^{-y_2},\bf{0_T})$
are introduced into the un-subtracted wave function. Compared to
Eq.~(\ref{TMD def: Collins}), the vector $u$ in the first (second)
piece of Wilson links $W_u$ ($W^{\dag}_{u}$) has been rotated slightly into the
space-like (time-like) direction $v$ ($n_2$) with large $-y_2$.
The orthogonality $n_2\cdot v=0$ implies that
the contribution of Fig~\ref{fig:quark-wilson-diagram}(f) vanishes in Feynman gauge,
and that a soft subtraction factor is not required in this definition.
That is, Eq.~(\ref{TMD def: Without soft factor}) will not cause double counting of soft gluons,
when it is implemented into a process more complicated than the pion transition
form factor, which demands soft-gluon factorization.

Computing all the one-loop graphs in Fig.~\ref{fig:quark-wilson-diagram}
according to Eq.~(\ref{TMD def: Without soft factor}), we derive
\begin{eqnarray}
{\phi}^{W(1)}_a (k_{+}^{\prime},k_T^{\prime},y_2)&=& {\alpha_s \, C_F \over 4 \, \pi} \,
\bigg [  \ln^2 \left ({ 2 \, e^{-2 y_2} \bar{k}_{+}^{2} \over  k_{T}^{2} } \right )  
-2 \, \ln \left ({ 2 \, e^{-2 y_2} \bar{k}_{+}^{2} \over  k_{T}^{2} } \right )  \bigg ]  
\delta(k_{+}-k_{+}^{\prime}) \, \delta ({\bf k}_T- {\bf k}_T^{\prime}) \, \nonumber \\
&& +{\alpha_s \, C_F \over \pi^2} \,
\bigg \{  \frac{\theta(k_{+}^{\prime}-k_{+})\, \theta(\bar k_{+}^{\prime}) }
{{( {\bf k}_T^{\prime}-{\bf k}_T)}^2 - \left ( {k_{+}-k_{+}^{\prime} \over p_{+}-k_{+}^{\prime} } \right ) {\bf k}_T^{\prime \,\, 2}}
\frac{ e^{-2 y_2} \, (k_{+} -k_{+}^{\prime}) }
{{( {\bf k}_T^{\prime}-{\bf k}_T)}^2 - 2 \, e^{-2 y_2} \, (k_{+}^{\prime} - k_{+})^2 }  \bigg \}_{\oplus} \,,
\label{result: 2a without soft factor} \nonumber \\
{\phi}^{W(1)}_b (k_{+}^{\prime},k_T^{\prime},y_2)&=& {\alpha_s \, C_F \over 4 \, \pi} \,
\bigg [  \ln^2 \left ({ 2 \, e^{-2 y_2} k_{+}^{2} \over  k_{T}^{2} } \right )  
-2 \, \ln \left ({ 2 \, e^{-2 y_2} k_{+}^{2} \over  k_{T}^{2} } \right ) + \pi^2  \bigg ]  
\delta(k_{+}-k_{+}^{\prime}) \, \delta ({\bf k}_T- {\bf k}_T^{\prime}) \, \nonumber \\
&& -{\alpha_s \, C_F \over \pi^2} \,
\bigg \{  \frac{\theta({k}_{+}^{\prime})\, \theta(k_{+}-k_{+}^{\prime}) }
{{( {\bf k}_T^{\prime}-{\bf k}_T)}^2 - \left ( {k_{+}^{\prime}-k_{+} \over k_{+}^{\prime} } \right ) {\bf k}_T^{\prime \,\, 2}}
\frac{ e^{-2 y_2} \, (k_{+}^{\prime} -k_{+}) }
{{({\bf k}_T^{\prime}-{\bf k}_T)}^2 + 2 \, e^{-2 y_2} \, (k_{+}^{\prime} - k_{+})^2 }  \bigg \}_{\oplus} \,,
\label{result: 2b  with Li definition}\nonumber  \\
{\phi}^{W(1)}_c(k_{+}^{\prime},k_T^{\prime},y_2)&=&\phi^{C(1)}_c(k_{+}^{\prime},k_T^{\prime})\,,
\nonumber 
\end{eqnarray}
\begin{eqnarray}
{\phi}^{W(1)}_d (k_{+}^{\prime},k_T^{\prime},y_2) &=& {\alpha_s \, C_F \over 4 \, \pi} \,
\bigg [ {1 \over \hat \epsilon} + \ln \left ( {\mu^2 \over k_T^{2}} \right )    
- \ln^2 \left ({ 2 \, e^{-2 y_2} k_{+}^{2} \over  k_{T}^{2} } \right )
+  \ln \left ({ 2 \, e^{-2 y_2} k_{+}^{2} \over  k_{T}^{2} } \right )
- {\pi^2 \over 3} + 2\bigg ]  \nonumber \\
&& \times \delta(k_{+}-k_{+}^{\prime}) \, \delta ({\bf k}_T- {\bf k}_T^{\prime}) \,,\nonumber\\
{\phi}^{W(1)}_e (k_{+}^{\prime},k_T^{\prime},y_2)
&=& {\phi}^{W(1)}_d (k_{+}^{\prime},k_T^{\prime},y_2) \big |_{k_{+} \to \bar{k}_{+}}  
- \pi^2 \, \delta(k_{+}-k_{+}^{\prime}) \, \delta ({\bf k}_T- {\bf k}_T^{\prime}) \,.
\end{eqnarray}
It is trivial to confirm that the sum of all the graphs in
Fig.~\ref{fig:quark-wilson-diagram} reproduces the $\ln k_T^2$ term
the same as in Eq.~(\ref{substraction: Collins}),
namely, the same as in \cite{Nandi:2007qx}.

\section{Equivalence of TMD Definitions}
\label{equiv}

We first point out that the TMD wave functions in Eqs.~(\ref{TMD def: Collins}) and
(\ref{TMD def: Without soft factor}) approach to the naive definition
\begin{eqnarray}
\phi^N(k_{+}^{\prime},k_T^{\prime},y_2)&=& \lim_{y_u \to - \infty}  \,\,
\int \frac{d z_{-}}{2 \pi} \int \frac{d^2 z_{T}}{(2 \pi)^2}
\,  e^{i(k_{+}^{\prime} z_{-} -k_T^{\prime} z_T)} \, \nonumber \\
&& \times  \langle 0| \bar d(0) W^{\dag}_{u}(+\infty,0)  \! \not  n_{-} \, \gamma_5 \,
W_{u}(+\infty,z) \, u(z) |\pi^{+}(p)  \rangle \,,
\label{naiveTMD}
\end{eqnarray}
in the limit of vanishing infrared regulators. It is easy
to see $S(z;y_2,y_u)= 1$ following Eq.~(\ref{soft}) and $S(z; y_1, y_2)= S(z;y_1,y_u)$
for the rapidities $y_2= y_u$, so that Eq.~(\ref{TMD def: Collins})
reduces to Eq.~(\ref{naiveTMD}) as $y_2= y_u\to -\infty$.
In the same limit both the gauge vectors $n_2$ and $v$
approach to $u$, and Eq.~(\ref{TMD def: Without soft factor}) also reduces to
Eq.~(\ref{naiveTMD}). The infinitesimal components $v^+ = -e^{y_u}$ and $u^+ = e^{y_u}$,
 being opposite in sign, serve as regulators for the rapidity divergences.
It has been known that the regularization of rapidity divergences,
which do not exist in QCD diagrams, is a matter
of factorization schemes \cite{Co03}. That is, Eq.~(\ref{TMD def: Without soft factor})
collects the same collinear divergences as Eq.~(\ref{TMD def: Collins}) which
are associated with the initial pion in the limit $y_2= y_u\to -\infty$.

We then demonstrate that Eqs.~(\ref{TMD def: Collins}) and (\ref{TMD def: Without soft factor})
collect the same collinear divergences for arbitrary rapidity $y_2$ as well.
The TMD wave function in Eq.~(\ref{TMD def: Collins}) depends on the Lorentz invariants
$p\cdot n_2$, $n_2^2$, and $k^2=-k_T^2$ formed by the vectors $p$, $k$ and
$n_2$. An infrared divergence is regularized by the parton virtuality $k^2$
into $\ln k_T^2$ in $k_T$ factorization as indicated by the one-loop result
in Eq.~(\ref{substraction: Collins}). Because the argument of a logarithm is dimensionless, $k_T$
appears in the ratio $p_+^2/k_T^2$ or $\mu^2/k_T^2$. Equations~(\ref{TMD def: Collins}) and
(\ref{TMD def: Without soft factor}) contain the same infrared logarithm
$\ln(\mu^2/k_T^2)$, which is generated by a loop correction without involving the Wilson links.
Therefore, we just focus
on the logarithm $\ln(p_+^2/k_T^2)$ in the two TMD definitions. Since the
Feynman rule $n_2^\mu/n_2\cdot l$ associated with the Wilson
link is scale invariant in $n_2$, $p_+^2/k_T^2$ must arise from the ratio
$(p\cdot n_2)^2/(n_2^2k^2)\propto (p_+^2e^{-2y_2})/k_T^2$ for Eq.~(\ref{TMD def: Collins}).
Equation~(\ref{TMD def: Without soft factor}) depends on the additional vector $v$
but with $n_2\cdot v=0$. The arguments of its infrared logarithms
are then given by $(p\cdot n_2)^2/(n_2^2k^2)$ and $(p\cdot v)^2/(v^2k^2)$, which are
both proportional to $(p_+^2e^{-2y_2})/k_T^2$. To study the infrared behaviors of
Eqs.~(\ref{TMD def: Collins}) and (\ref{TMD def: Without soft factor}) for arbitrary $y_2$,
we vary $y_2$ below.

Consider the derivative
\begin{eqnarray}
\frac{d}{dy_2}\phi^C=\frac{n_2^2}{2p\cdot n_2}p^\alpha\frac{d}{dn_2^\alpha}\phi^C\,,
\label{d1}
\end{eqnarray}
which is a straightforward consequence of the chain rule \cite{CS81}.
The differentiation $d/dn_2^\alpha$ applies to the Wilson links in the direction of
$n_2$, leading to the Feynman rule
\begin{eqnarray}
\frac{n_2^2}{2p\cdot n_2}p^\alpha\frac{d}{dn_2^\alpha}\frac{n_2^\mu}{n_2\cdot l}=
\frac{\hat n_2^\mu}{n_2\cdot l}\,,
\end{eqnarray}
with the special vertex \cite{L97}
\begin{eqnarray}
\hat n_2^\mu=\frac{n_2^2}{2p\cdot n_2}\left(p^\mu-\frac{p\cdot l}{n_2\cdot l}n_2^\mu\right)\,.
\label{n2}
\end{eqnarray}
Equation~(\ref{d1}) then yields
\begin{eqnarray}
\frac{d}{dy_2}\phi^C= \lim_{\substack{y_1 \to + \infty  \\   y_u \to - \infty}}  \,
\frac{1}{2}\left[\frac{S^{\,\prime}(z; y_1, y_2)}{S(z; y_1, y_2)}-
\frac{S^{\,\prime}(z;y_2,y_u)}{S(z;y_2,y_u)}\right]\phi^C\,,\label{d11}
\end{eqnarray}
in coordinate space, where the primed soft functions $S^{\,\prime}$ include the diagrams
from those in the soft functions $S$, with an original vertex $n_2^\mu$ being replaced by
a special vertex $\hat n_2^\mu$ on the Wilson links in the direction of $n_2$.

In the leading-power approximation, the accuracy at which Eq.~(\ref{TMD def: Collins})
is defined, the diagrams in $S^{\,\prime}(z; y_1, y_2)$ are
organized into a product of the soft function $S(z; y_1, y_2)$ with
a soft kernel $K(z; y_1, y_2)$ following the argument in \cite{Li:2013xna,L97}.
The soft kernel $K(z; y_1, y_2)$ contains the same set of diagrams as the soft
function $S(z; y_1, y_2)$ at each order of the strong coupling constant, but with a
special vertex on the Wilson links
in the direction of $n_2$ \cite{Li:2013xna,L97}.
Similarly, $S^{\,\prime}(z;y_2,y_u)$ is expressed as a product of $S(z;y_2,y_u)$ and
$K(z;y_2,y_u)$ at leading power, so Eq.~(\ref{d11}) is simplified into
\begin{eqnarray}
\frac{d}{dy_2}\phi^C= \lim_{\substack{y_1 \to + \infty  \\   y_u \to - \infty}}  \,
\frac{1}{2}\left[K(z; y_1, y_2)-K(z;y_2,y_u)
\right]\phi^C\,.\label{d12}
\end{eqnarray}
Because the special vertex suppresses collinear dynamics \cite{Li:2013xna,L97}, the soft
kernels $K(z; y_1, y_2)$ and $K(z;y_2,y_u)$ collect only the single logarithms
$\ln(n_2\cdot n_1)$ and $\ln(n_2\cdot u)$, respectively.
With the relation between the infrared logarithms in the limit
$y_1=-y_u\to\infty$, $\ln(n_2\cdot u)=-\ln(n_2\cdot n_1)$,
which holds for arbitrary finite $y_2$, we have $K(z; y_1, y_2)\approx -K(z;y_2,y_u)$
up to different infrared finite pieces, and
\begin{eqnarray}
\frac{d}{dy_2}\phi^C\approx  \lim_{y_1 \to + \infty}  \,
K(z; y_1, y_2)\phi^C\,,\label{d13}
\end{eqnarray}
from Eq.~(\ref{d12}).

Both the variations with respect to $n_2$ and $v$ are related to
the variation of $y_2$ via the chain rule, so the above derivation applies
to the TMD wave function in Eq.~(\ref{TMD def: Without soft factor}). We obtain
\begin{eqnarray}
\frac{d}{dy_2}\phi^W\equiv \left[\frac{n_2^2}{2p\cdot n_2}p^\alpha\frac{d}{dn_2^\alpha}
+\frac{v^2}{2p\cdot v}p^\alpha\frac{d}{dv^\alpha}\right]\phi^W,
\label{d2}
\end{eqnarray}
where the first (second) term on the right hand side includes the diagrams
from those in $\phi^W$, with an original vertex $n_2^\mu$ ($v^\mu$) being replaced by
a special vertex $\hat n_2^\mu$ ($\hat v^\mu$) on the Wilson
link in the direction of $n_2$ ($v$). The definition of the special vertex $\hat v^\mu$
is similar to $\hat n_2^\mu$ in Eq.~(\ref{n2}) but with the vector $n_2$ being replaced
by $v$. A subset of diagrams, in which the gluon emitted by the special vertex $\hat n_2^\mu$
($\hat v^\mu$) carries a small momentum, is factorized out of the first (second) derivative
in Eq.~(\ref{d2}). The resultant soft kernel is composed of a pair of Wilson links in the
direction of $n_1$, which are collimated to the initial quarks in the limit $y_1\to\infty$
\cite{Li:2013xna,L97}, a Wilson link in the direction of $n_2$, and a Wilson link in the
direction of $v$. A special vertex $\hat n_2^\mu$ ($\hat v^\mu$) appears on the Wilson
link in the direction of $n_2$ ($v$) in the first (second) soft kernel. Due to the
suppression from the special vertices on collinear dynamics, the first and second
soft kernels collect only the single logarithms $\ln(n_2\cdot n_1)$ and
$\ln(v\cdot n_1)$, respectively. It is obvious that these two logarithms are equal in
the limit $y_1\to\infty$, and can be combined into the soft kernel $K(z; y_1, y_2)$.

Note that the Wilson links along $n_2$ and $v$ attach to the energetic quarks,
instead of to other Wilson links.
In addition to the soft kernel $K$ factorized above, another subset of diagrams,
in which the gluon emitted by the special vertex and attaching to the quark line
carries a large (but not collinear) momentum, can also be factorized \cite{CS81}.
This factorization follows the argument in \cite{Li:2013xna}, and the resultant hard
kernel $G(z,y_2)$ contains a special vertex on the Wilson link in the direction of $n_2$ or
of $v$. Hence, the two terms in Eq.~(\ref{d2}) are summed into the product
\begin{eqnarray}
\frac{d}{dy_2}\phi^W=\lim_{y_1 \to + \infty}  \,
[K(z; y_1, y_2)+G(z,y_2)]\phi^W.\label{d23}
\end{eqnarray}
The functions $K$ and $G$ correspond to the known soft and hard kernels in the
typical Sudakov resummation \cite{CS81}, both of which can be evaluated order by order
according to their definitions described above, with their one-loop expressions being
found in \cite{Li:2013xna}. We have confirmed that the
resultant rapidity evolution equation is
the same as the one derived in \cite{Nandi:2007qx} in the small $k_+^{\prime}$ limit, namely,
in the so-called small $x$ limit, where the $k_T$ factorization theorem is an
appropriate theoretical framework for exclusive processes.
Note that $K$ and $G$ depend on a factorization scale $\mu$,
which cancels in their sum $K+G$. The $\mu$-dependent kernel in Eq.~(\ref{d13}) was
also observed in the rapidity evolution kernel for the TMD fragmentation function
(see Eq. (13.55) of \cite{Collins:2011zzd}), and calls for a simultaneous
treatment of the rapidity and factorization-scale evolutions.

We have shown that $\phi^C$ and $\phi^W$ reduce to the naive TMD wave function
as $y_2=y_u\to-\infty$.
Apparently, the hard kernel $G$ does not depend on the infrared logarithm $\ln k_T$,
and can be regarded as a finite piece. Equations~(\ref{d13}) and (\ref{d23}), governed
by the identical soft kernel $K$, then imply that
$\phi^C$ and $\phi^W$ have the same infrared logarithms at leading power
for arbitrary $y_2$. However, they are established in different factorization
schemes represented by the infrared finite piece $G$. We claim that the two
TMD definitions considered in this work are equivalent
in the infrared behavior at all orders of the strong coupling constant,
and supersede the one presented in \cite{Li:2013xna}.

\section{Conclusion}
\label{section: summary}

In this paper we have first investigated the infrared behavior
of a TMD pion wave function with the dipolar Wilson
links and the complicated soft subtraction, which was originally developed for a
TMD parton density. The TMD wave-function definition with non-dipolar
off-light-cone Wilson links was then proposed, which was shown to realize the
$k_T$ factorization of hard exclusive processes appropriately as well.
It is free of the rapidity divergence and of the pinched singularity
in the self-energy correction to the dipolar Wilson lines, and demands
simpler soft subtraction. We have illustrated its property by considering the
special case with two orthogonal gauge vectors, for which the soft subtraction is
not needed in Feynman gauge. It was explicitly demonstrated at one-loop
level that this definition yields the collinear logarithms $\ln k_T^2$ the same
as in the one with the dipolar gauge links, which
cancel those in the QCD diagrams, albeit with a distinct ultraviolet
structure. We then illustrated the equivalence of the two definitions
by showing that both of them reduce to the naive TMD wave function as the
non-light-like Wilson links approach to the light cone, and that their
evolutions with the rapidity of the non-light-like Wilson links are
governed by the same soft kernel. In this reasoning it also became clear that
the two TMD wave functions were established in different factorization schemes.

As stressed at the beginning of Sec.~\ref{section: non-dipolar Wilson lines},
we started with the un-subtracted TMD wave function in Eq.~(\ref{TMD def: Collins}),
where the future-pointing or past-pointing light-like Wilson links have been
appropriately chosen to facilitate the $k_T$ factorization by avoiding the Glauber region.
Therefore, our proposal for a TMD wave function facilitates proofs of the $k_T$ factorization
theorem for hard exclusive reactions, and derivations of their various evolution
equations.
It is then crucial to explore phenomenological consequences of applying
the new TMD definition, which includes evolution effects, to $k_T$ factorization formulas
for exclusive processes. It is straightforward to extend our
proposal to the definition of the $B$ meson TMD wave functions in the heavy-quark effective
theory, which will put the perturbative QCD factorization approach to exclusive $B$ meson
decays on more solid ground. It is also of interest to examine the impact of the new
TMD definition on polarized processes, for which Wilson-link interactions play an
important role. We plan to study the above topics in future publications.

\section*{Acknowledgement}

We thank John Collins and Zhongbo Kang for illuminating discussions.
HNL is supported  in part by the Ministry of Science and Technology of R.O.C. under
Grant No. NSC-101-2112-M-001-006-MY3.  YMW acknowledges support by the
{\it DFG-Sonder\-forschungs\-bereich/Trans\-regio 9 ``Computer\-gest\"{u}tzte
Theoretische Teilchen\-physik"}.

%
%

\end{document}